\documentclass[prb,superscriptaddress,twocolumn,showpacs,floatfix,aps,10pt]{revtex4-1}
\usepackage{amsmath}
\usepackage{amssymb}
\usepackage{graphicx}
\usepackage{natbib}
\usepackage{subfigure}
\usepackage[colorlinks=true, linkcolor=blue, citecolor=blue, 
filecolor=blue, urlcolor=blue, bookmarks=true, bookmarksopen=true, 
bookmarksopenlevel=3, plainpages=false, pdfpagelabels=true]{hyperref}

\begin{document}
\title{Derivation of phenomenological expressions for transition matrix elements for electron-phonon scattering}
\author{C.~Illg}
\author{M.~Haag}
\author{B.Y.~Mueller}
\affiliation{Max Planck Institute for Intelligent Systems, Heisenbergstr.\ 3, 70569 Stuttgart, Germany}
\author{G.~Czycholl}
\affiliation{Institut f\"ur theoretische Physik, Universit\"at Bremen, Bibliothekstr. 1, 28359 Bremen, Germany}
\author{M.~F\"ahnle}
\email[Electronic address: ]{faehnle@is.mpg.de}
\affiliation{Max Planck Institute for Intelligent Systems, Heisenbergstr.\ 3, 70569 Stuttgart, Germany}

\begin{abstract}
In the literature on electron-phonon scatterings very often a phenomenological expression
for the transition matrix element is used which was derived in the textbooks of Ashcroft/Mermin and
of Czycholl. There are various steps in the derivation of this expression.
In the textbooks in part different arguments have been used in these steps, 
but the final result is the same.
In the present paper again slightly different arguments are used
which motivate the procedure in a more intuitive way.
Furthermore, we generalize the phenomenological expression to describe the dependence of the 
matrix elements on the spin state of the initial and final electron state. 

\end{abstract}
\maketitle

\section{Introduction}\label{secI}
Many phenomena in solid state physics are determined by the scattering of electrons at phonons.
Examples are the electrical and thermal
conductivity, the lifetime of excited 
electron states~\cite{Knorren2001}
and a possible contribution
to the ultrafast demagnetization
after irradiation of a ferromagnetic film
by a femtosecond laser pulse (see below).
For a monatomic crystal the transition rate $W^{\lambda}_{j\mathbf{k},j'\mathbf{k'}}$ for a transition from an electronic state $\left|j\mathbf{k}\right>$ with energy $\epsilon_{j\mathbf{k}}$ 
to a state $\left|j'\mathbf{k}'\right>$ with energy $\epsilon_{j'\mathbf{k}'}$ ($j$,$j'$ denote the band indices, $\mathbf{k}$, $\mathbf{k}'$ the wave vectors and $\lambda$ denotes the three polarization vectors $\mathbf{p}_{\mathbf{q}\lambda}$ of a phonon)
is given by Fermi's golden rule,
\begin{align}\label{gl1}
&W^{\lambda}_{j\mathbf{k},j'\mathbf{k'}}=\frac{2\pi}{\hbar}\left|M^{\lambda}_{j\mathbf{k},j'\mathbf{k'}}\right|^{2}\\ \notag
&\left\{n_{\mathbf{q}\lambda}\delta\left[\epsilon_{j'\mathbf{k'}}-\left(\epsilon_{j\mathbf{k}}+\hbar\omega_{\mathbf{q}\lambda}\right)\right]+\right.\\ \notag
&\left.\left(n_{\mathbf{-q}\lambda}+1\right)\delta\left[\epsilon_{j'\mathbf{k'}}-\left(\epsilon_{j\mathbf{k}}-\hbar\omega_{-\mathbf{q}\lambda}\right)\right] \right\}.
\end{align}
The phonon with wavevector $\mathbf{q}$ has the frequencies $\omega_{\mathbf{q}\lambda}$,  $n_{\mathbf{q}\lambda}=\left[\exp\left(\hbar\omega_{\mathbf{q}\lambda}/\text{k}_{\text{B}}T_{\text{p}}\right)-1\right]^{-1}$ is the Bose distribution function with the phonon temperature $T_{\text{p}}$.
Absorption of phonons and both induced and spontaneous emission of phonons are included in Eq.\ \eqref{gl1}.
$M^{\lambda}_{j\mathbf{k},j'\mathbf{k'}}$ is the transition matrix element
\begin{align}\label{gl2}
M^{\lambda}_{j\mathbf{k},j'\mathbf{k'}}=\left<\Psi_{j'\mathbf{k'}}\left|V_{\text{ph}}\right|\Psi_{j\mathbf{k}}\right>,
\end{align}
where $V_{\text{ph}}$ is the electron-phonon scattering operator and $\mathbf{q}=\pm(\mathbf{k'}-\mathbf{k}+\mathbf{G})$ where the $+(-)$ sign holds for phonon absorption (emission) and where $\mathbf{G}$ is a reciprocal lattice vector which brings $\mathbf{k'}$ back to the first Brillouin zone  if $\mathbf{k}'=\mathbf{k}+\mathbf{q}$ is  outside. 
Eq.\ \eqref{gl2} includes spin-flip transitions although the spin does not explicitly occur in the wavefunctions $\Psi_{j\mathbf{k}}$.
The reason is that the dominant spin character of the state is determined by j and $\mathbf{k}$,
and it may  well be different for $\Psi_{j\mathbf{k}}$ and $\Psi_{j'\mathbf{k}'}$.
The  operator $V_{\text{ph}}$ has two contributions \cite{1}, one resulting from the phonon-induced distortion of the lattice potential 
(which is often called Elliott part, standard
part or Fr\"ohlich part) and one resulting from the  phonon-induced distortion of the spin-orbit coupling (Yafet part).
In the Elliott-part only the phonon-induced modifications of the spin-diagonal part of the effective potential which the electrons feel enters,
which is called deformation potential.
Elliott has shown \cite{Elliott} that this spin-diagonal part nevertheless leads to spin-flip scatterings
in systems with spin-orbit  coupling.
Yafet has shown \cite{1} that one must take into account both contributions to get the correct $q=\left|\mathbf{q}\right|\rightarrow 0$ limit
for the behavior of the matrix element, 
and this has been confirmed by Grimaldi and Fulde \cite{5}.

It is well known that the modification of the lattice potential and of the spin-orbit coupling
due to the phonon is changed by the reaction of the electrons (mainly the valence electrons)
and the ions on the displacement of the ions by the phonon, i.e.,
there is electronic and ionic screening of the electron-phonon
transition matrix element, especially for small $\left|\mathbf{q}\right|$.
The screening of the Elliott part has been discussed by Kittel \cite{Kittel} within
a complicated perturbation theory, and in the book of Giuliani and Vignale \cite{Vignale}
with Green function based methods.
A simplified version of a phenomenological screening theory is given in the textbooks of
Ashcroft and Mermin \cite{7} and  of Czycholl \cite{8}.
In this simplified version it is assumed (see section \ref{secII}) that the matrix
element (which is in general $M^{\lambda}_{j\mathbf{k},j'\mathbf{k}'}$, see above)
depends only on $\mathbf{q}$ and on the phonon frequency
$\omega_{\mathbf{q}}$, yielding the approximation
\begin{align}
 \label{gl3n}
\left|M\left(\mathbf{q}\right)\right|^{2}=\frac{4\pi e^{2}}{V\left(q^{2}+k_{\text{TF}}^{2}\right)}\frac{1}{2}\hbar\omega_{\mathbf{q}}.
\end{align}
Here $V$ is the volume of the system, and $k_{\text{TF}}$ is the Thomas-Fermi wavevector of the system.
Because Eq.\ \eqref{gl3n} does not contain the spin it gives the same result independent of the question whether the
considered electron-phonon scattering is a spin-flip transition or a non-spin-flip transition.
This is corrected in Sec.\ \ref{secIII} by multiplying $\left|M\left(\mathbf{q}\right)\right|^{2}$ from Eq.\ \eqref{gl3n}
with $4b^{2}$ where $b$ is the spin-mixing factor.
This resulting Elliott part therefore can describe spin-flip transitions.

The screening of the Yafet part has been discussed for paramagnetic systems by Yafet \cite{1} 
using an a-priori screened lattice potential, and by Grimaldi and Fulde \cite{5} within
a self consistent Hartree approximation.
For magnetic systems Rajagopal and Mochena \cite{6} have discussed screening effects beyond the theory of Grimaldi and Fulde
for paramagnetic systems.
However, no explicit results for the screened electron-phonon matrix element have been given which 
could be used in electron-theoretical treatments of electron-phonon scattering.
To get a simple phenomenological approximation for the screened matrix elements
$M^{\lambda}_{j\mathbf{k},j'\mathbf{k}'}$ which can be used in such treatments 
one must try to find explicit results of the theory of Ref.\ \onlinecite{6} 
and to approximate them as strongly as possible.
In the literature on electron-theoretical treatments of electron-phonon
scattering the matrix elements $M^{\lambda}_{j\mathbf{k},j'\mathbf{k}}$ have
been calculated on the one hand by the ab-initio density functional electron theory
and on the other hand by using phenomenological approximations based on Eq.\ \eqref{gl3n}.

\subsection*{phenomenological approximations}

There are various steps in the derivation of Eq.\ \eqref{gl3n}.
In Ref.\ \onlinecite{7} and \onlinecite{8} in part different arguments have been used in these steps,
but the final result is the same.
In Sec.\ \ref{secII} we follow mainly the approach of Ref.\ \onlinecite{8},
but again we use at least slightly different arguments which in our opinion motivate the procedure 
in a more  intuitive way.
Because the screening theory in principle is very complicated, we think that it is important to get
an intuitive feeling for the arguments of the phenomenological screening theory.
Our Sec.\ \ref{secII} leads to the same result, Eq.\ \eqref{gl3n}, as published already in textbooks \cite{7,8}.
Therefore one could argue that there is no novelty in the whole Sec.\ \ref{secII}.
However, the  novelty is a better intuitive understanding.
A second reason to describe the (slightly modified) approach of Ref.\ \onlinecite{8} 
in detail is that Ref.\ \onlinecite{8} is a textbook in German language which
may not be readily available for many readers.

\section{Derivation of the approximated electron-phonon transition matrix element}\label{secII}
\subsection{Phenomenological theory of screening of the Coulomb interaction between two electrons}\label{secIIa}
In the phenomenological screening theory \cite{7,8} the screening of the Coulomb interaction matrix element (Eq.\ (5.42) of Ref.\ \onlinecite{8})
\begin{align}\label{gl8}
 V_{\mathbf{k},\mathbf{k}'}=\frac{1}{V}\frac{4\pi e^{2}}{q^{2}}
\end{align}
for the Coulomb interaction between two free electrons with charge $e$ and with wave vectors $\mathbf{k}$ and $\mathbf{k}'$ ($q=\left|\mathbf{q}\right|=\left|\mathbf{k}-\mathbf{k}'\right|$) is calculated ($V$ is the volume of the crystal). To do this, the Coulomb potential between these two electrons is considered as being an "external" potential.
The screened form of $V_{\mathbf{k}\mathbf{k}'}$ then is
\begin{align}\label{gl9}
 V^{\text{eff}}_{\mathbf{k},\mathbf{k}'}=\frac{1}{V}\frac{4\pi e^{2}}{\epsilon\left(\mathbf{q}\right)q^{2}},
\end{align}
with the dielectric function which is given by
\begin{align}\label{gl10}
 \frac{1}{\epsilon(\mathbf{q})}=\frac{1}{\epsilon_{\text{electronic}}(\mathbf{q})}\frac{1}{\epsilon_{\text{ionic}}(\mathbf{q})}.
\end{align}
For free electrons $\epsilon_{\text{electronic}}\left(\mathbf{q}\right)$ is given by the Thomas-Fermi screening,
\begin{align}\label{gl11}
 \epsilon_{\text{TF}}=\frac{q^{2}+k^{2}_{\text{TF}}}{q^{2}}.
\end{align}
The quantity $\epsilon_{\text{ionic}}\left(\mathbf{q}\right)$ is given by \cite{7,8}
\begin{align}\label{gl12}
\epsilon_{\text{ionic}}\left(\mathbf{q}\right)=\frac{\omega^{2}-\omega^{2}(\mathbf{q})}{\omega^{2}},
\end{align}
where $\hbar\omega=\epsilon\left(\mathbf{k}\right)-\epsilon\left(\mathbf{k'}\right)$ with the electronic energies $\epsilon\left(\mathbf{k}\right)$, $\epsilon\left(\mathbf{k}'\right)$, and where $\omega(\mathbf{q})$ is the frequency of a phonon with wave vector $\mathbf{q}$, whereby the dependence on $\lambda$ is neglected.
We note a slight inconsistency in the (accepted) phenomenological derivation 
of Eq.\ \eqref{gl12}.
To derive this equation, one starts with the notion that all ions are uniformly displaced from their equilibrium positions.
If the electrons remain in their equilibrium positions, then the ions would perform a collective plasma oscillation with an ionic plasma  frequency \cite{8} $\omega_{\text{p}}$. However, the electrons react on the uniform motion of the ions and screen the thereby generated polarization.
This electronic screening is described by Eq.\ \eqref{gl11}.
Due to this  screening the electric field produced by the ion displacement is a factor of $\frac{1}{\epsilon_{\text{TF}}(\mathbf{q})}$ smaller than the unscreened field.
Therefore the plasma frequency $\omega_{\text{p}}$ is modified by the $\mathbf{q}$-dependent electronic dielectric function, leading to $\mathbf{q}$-dependent acoustic phonon frequencies 
\cite{8}
$\omega^2(\mathbf{q})=\omega^2_{\text{p}}/\epsilon_{\text{electronic}}(\mathbf{q})$.
It is a bit inconsistent to start with a uniform plasma motion of the ions and generate a $\mathbf{q}$-dependent phonon frequency by introducing the $\mathbf{q}$-dependent electronic screening, 
although the primary 
motion of the ions (which is screened) is not $\mathbf{q}$-dependent.

The electrons in states $\left|\mathbf{k}\right>$ and $\left|\mathbf{k}'\right>$ experience a time-dependent Coulomb potential.
Since the mass of the ion is larger than the electronic mass, the ions cannot react instantaneously on this potential, and therefore the  ionic dielectric function depends on the one hand on the phonon frequency $\omega\left(\mathbf{q}\right)$ and on the other hand on a frequency which characterizes the electronic situation.
It is  physically meaningful to insert for it $\omega=\frac{\epsilon\left(\mathbf{k}\right)-\epsilon\left(\mathbf{k'}\right)}{\hbar}$, although there is no strict mathematical proof for this.
It is  a bit inconsistent to start with the  time-independent perturbation given by Eq.\ \eqref{gl8} and to introduce then a frequency-dependent screening.
In principle, these problems could be avoided by discussing from the beginning a time-dependent treatment as  in the book
of Giuliani and Vignale \cite{Vignale}, but  these treatments did not lead to simple phenomenological expressions.

If we had only electronic screening, then we would have
\begin{align}\label{gl13}
 V^{\text{eff},\text{el}}_{\mathbf{k},\mathbf{k}'}=\frac{1}{V}\frac{4\pi e^{2}}{q^{2}+k_{\text{TF}}^{2}},
\end{align}
i.e., the additional term
\begin{align}\label{gl14}
 V^{\text{el}}_{\mathbf{k},\mathbf{k}'}=\frac{1}{V}\frac{4\pi e^{2}}{q^{2}}\cdot\left(\frac{1}{1+\frac{k^{2}_{\text{TF}}}{q^{2}}}-1\right),
\end{align}
to $V_{\mathbf{k},\mathbf{k}'}$ of Eq.\ \eqref{gl8}. With the ionic screening we get a second additional term,
\begin{align}\label{gl15}
 V^{\text{eff},\text{ion}}_{\mathbf{k},\mathbf{k}'}=\frac{1}{V}\frac{4\pi e^{2}}{q^{2}+k_{\text{TF}}^{2}}\frac{\omega^{2}\left(\mathbf{q}\right)}{\omega^{2}-\omega^{2}\left(\mathbf{q}\right)}.
\end{align}
Altogether, the matrix element of the screened Coulomb interaction is
\begin{align}\label{gl16}
 V^{\text{eff}}_{\mathbf{k},\mathbf{k}'}=&\frac{1}{V}\left[\frac{4\pi e^{2}}{q^{2}}+\frac{4\pi e^{2}}{q^{2}}\cdot\left(\frac{1}{1+\frac{k^{2}_{\text{TF}}}{q^{2}}}-1\right)\right.+\\ \notag
                                  &\left.\frac{4\pi e^{2}}{q^{2}+k_{\text{TF}}^{2}}\frac{\omega^{2}\left(\mathbf{q}\right)}{\omega^{2}-\omega^{2}\left(\mathbf{q}\right)}\right].
\end{align}
This gives:
\begin{align}\label{gl17}
 V^{\text{eff}}_{\mathbf{k},\mathbf{k}'}=\frac{1}{V}\frac{4\pi e^{2}}{q^{2}+k_{\text{TF}}^{2}}\left[1+\frac{\omega^{2}\left(\mathbf{q}\right)}{\omega^{2}-\omega^{2}\left(\mathbf{q}\right)}\right],
\end{align}
which is Eq.\ \eqref{gl9} with Eqs.\ \eqref{gl11}, \eqref{gl12}. 
We have discussed this in detail because in sec.\ \ref{secIIc} we represent the term $V^{\text{ion}}_{\mathbf{k}\mathbf{k}'}$ by a special matrix element which we calculate by perturbation theory.
Thereby we have to omit one of the terms of the perturbation series which represents $\frac{4\pi e^{2}}{q^{2}}$ and terms which are related to purely electronic screening effects.

\subsection{Comments on the  dielectric function $\epsilon(\mathbf{q})$}\label{secIIb}
In Sec.\ \ref{secIIa} a simple approximation for the dielectric function $\epsilon\left(\mathbf{q}\right)$ has been used
to obtain the screened form of the effective Coulomb interaction between two free electrons. 
A better approximation is, e.g., the Lindhard dielectric constant,
and even more better but more complicated approximations are discussed in Kittel's book \cite{Kittel}.
In the present chapter we show that the dielectric function $\epsilon\left(\mathbf{q}\right)$ in  Eq.\ \eqref{gl9} 
is the same function as used under certain circumstances in electrodynamics to relate the displacement field $\mathbf{D}$
of a dielectric material  to the electric field in the material.
The reason is that if this equivalence holds then one can use ab-initio codes to determine $\epsilon\left(\mathbf{q}\right)$
very accurately and insert this  $\epsilon\left(\mathbf{q}\right)$ into Eq.\ \eqref{gl9} to get a better theory for $V^{\text{eff}}_{\mathbf{k},\mathbf{k}'}$.
In Ref.\ \onlinecite{7} it is assumed that this equivalence holds, we want to prove it.
It should be noted that in the literature it has been already shown by totally different methods
that this equivalence holds under certain circumstances, i.e. when local field corrections are neglected \cite{Sturm,Ehrenreich}.
Therefore the fact that this equivalence can be proved is no novelty, however,
the way we prove it is different from the old proofs, 
and we think that it is interesting to see that the same result can be obtained by seemingly completely different arguments.

As discussed in Sec.\ \ref{secIIa}, the Coulomb potential between electrons is considered as "external potential" $\Phi^{\text{ext}}$,
 i.e., the charges of these two electrons are considered as "external charges" with ($\delta$-shaped) charge density $\rho^{\text{ext}}\left(\mathbf{r}\right)$.
The bare potential $\Phi^{\text{ext}}$ of $\rho^{\text{ext}}$ follows from Poisson's  equation
\begin{align}\label{gl18}
 \Delta\Phi^{\text{ext}}\left(\mathbf{r}\right)=-4\pi\rho^{\text{ext}}\left(\mathbf{r}\right).
\end{align}
The screening of $\Phi^{\text{ext}}\left(\mathbf{r}\right)$ occurs by the reaction of the other electrons and the ions on $\rho^{\text{ext}}\left(\mathbf{r}\right)$, leading to an induced charge density $\rho^{\text{ind}}\left(\mathbf{r}\right)$, so that the total charge density $\rho^{\text{total}}\left(\mathbf{r}\right)$ is
\begin{align}\label{gl19}
 \rho^{\text{total}}\left(\mathbf{r}\right)=\rho^{\text{ext}}\left(\mathbf{r}\right)+\rho^{\text{ind}}\left(\mathbf{r}\right).
\end{align}
The  screened potential $\Phi^{\text{total}}\left(\mathbf{r}\right)$ is calculated from
\begin{align}\label{gl20}
 \Delta\Phi^{\text{total}}\left(\mathbf{r}\right)=-4\pi\rho^{\text{total}}(\mathbf{r}).
\end{align}
We consider a homogeneous system with a linear relation between $\Phi^{\text{total}}\left(\mathbf{r}\right)$ and $\Phi^{\text{ext}}\left(\mathbf{r}\right)$,
\begin{align}\label{gl21}
 \Phi^{\text{ext}}\left(\mathbf{r}\right)=\int\limits_{\text{sample}}\epsilon\left(\mathbf{r}-\mathbf{r'}\right)\Phi^{\text{total}}\left(\mathbf{r'}\right)d^{3}r'.
\end{align}
We now apply $\nabla_{\mathbf{r}}$ to Eq.\ \eqref{gl21}, yielding
\begin{align}\label{gl22}
 \nabla_{\mathbf{r}}\Phi^{\text{ext}}\left(\mathbf{r}\right)=&\int\limits_{\text{sample}}\nabla_{\mathbf{r}}\epsilon\left(\mathbf{r}-\mathbf{r}'\right)\Phi^{\text{total}}\left(\mathbf{r}'\right)d^{3}r'=\\ \notag
 &-\int\limits_{\text{sample}} \nabla_{\mathbf{r}'}\epsilon\left(\mathbf{r}-\mathbf{r}'\right)\Phi^{\text{total}}\left(\mathbf{r}'\right)d^{3}r'.
\end{align}
Performing a partial integration we get
\begin{align}\label{gl23}
&\int \nabla_{\mathbf{r}'}\epsilon\left(\mathbf{r}-\mathbf{r}'\right)\Phi^{\text{total}}\left(\mathbf{r}'\right)d^{3}r'=\\ \notag
&-\int\limits_{\text{sample}}\epsilon\left(\mathbf{r}-\mathbf{r}'\right)\nabla_{\mathbf{r}'}\Phi^{\text{total}}\left(\mathbf{r}'\right)d^{3}r'+\\ \notag&\oint\limits_{\text{surface of the sample}}\Phi^{\text{total}}\left(\mathbf{r}'\right)\epsilon\left(\mathbf{r}-\mathbf{r}'\right)\mathbf{n}\cdot\mathbf{dS},
\end{align}
where $\mathbf{n}\left(\mathbf{r}'\right)$ is the  local surface normal vector.

The  scope of our procedure is to investigate how an "external potential" $\Phi^{\text{ext}}$ in the interior of the material is screened.
We do not want to investigate how this screening depends on the distance between the "external charge" and the surface.
In fact we consider the external charge deep in the interior of a sample which is very large so that the surfaces are very far away from $\rho^{\text{ext}}$ and they do not have an influence on the screening of $\rho^{\text{ext}}$.
We therefore can consider a large sample of arbitrary shape, e.g., a sphere.
Far away from two electrons which are close to each other in the interior of a large sphere the charge density $\rho^{\text{ext}}$ looks spherically symmetric.
Thus $\Phi^{\text{total}}$ has the same value on the whole surface of the sphere.
Furthermore, $\epsilon\left(\mathbf{r}-\mathbf{r}'\right)$ is also the same on the whole surface of a sphere. For each $\mathbf{n}$ on one hemisphere there is an opposite $\mathbf{n}$
on the other hemisphere, and, altogether,  the surface integral in Eq.\ \eqref{gl23} is zero.
Using Eq.\ \eqref{gl23} this yields
\begin{align}\label{gl24E}
 \nabla_{\mathbf{r}}\Phi^{\text{ext}}\left(\mathbf{r}\right)=&\int\limits_{\text{sample}}d^{3}r'\epsilon\left(\mathbf{r}-\mathbf{r}'\right)\nabla_{\mathbf{r}'}\Phi^{\text{total}}\left(\mathbf{r}'\right).
\end{align}
Now we use the definitions  of $\mathbf{D}\left(\mathbf{r}\right)$ and $\mathbf{E}\left(\mathbf{r}\right)$ in electrodynamics,
\begin{align}\label{gl28E}
 \nabla_{\mathbf{r}}\Phi^{\text{ext}}\left(\mathbf{r}\right)=-\mathbf{D}\left(\mathbf{r}\right),\nabla_{\mathbf{r}}\Phi^{\text{total}}\left(\mathbf{r}\right)=-\mathbf{E}\left(\mathbf{r}\right)
\end{align}
to get $\mathbf{D}\left(\mathbf{r}\right)=\int\limits_{\text{sample}}\epsilon\left(\mathbf{r}-\mathbf{r}'\right)\mathbf{E}\left(\mathbf{r}'\right)d^{3}r'$, i.e., we have shown that the dielectric function occurring in Eq.\ \eqref{gl9} is indeed the same as the dielectric function of electrodynamics between $\mathbf{D}$ and $\mathbf{E}$.
The most general linear relation between $\mathbf{D}$ and $\mathbf{E}$ in a homogeneous but not isotropic system is
\begin{align}\label{gl29E}
 \mathbf{D}\left(\mathbf{r}\right)=\int\limits_{\text{sample}}\underline{\underline{\epsilon}}\left(\mathbf{r}-\mathbf{r}'\right)\mathbf{E}\left(\mathbf{r}'\right)d^{3}r'
\end{align}
with the dielectric tensor $\underline{\underline{\epsilon}}\left(\mathbf{r}-\mathbf{r}'\right)$. For an isotropic system $\underline{\underline{\epsilon}}$ has only diagonal components which are all the same, 
$\epsilon\left(\left|\mathbf{r}-\mathbf{r}'\right|\right)$.

\subsection{Representation of $V^{\text{ion}}_{\mathbf{k},\mathbf{k'}}$ by a transition matrix element for a many-free-electron system} \label{secIIc}

We denote the inverse Fourier transform of $V^{\text{ion}}_{\mathbf{k},\mathbf{k}'}$ of Eq.\ \eqref{gl15} as $V^{\text{ion}}\left(\mathbf{r}-\mathbf{r}'\right)$ so that
\begin{align}\label{gl24}
 &\frac{1}{V}\frac{4\pi e^{2}}{q^{2}+k_{\text{TF}}^{2}}\frac{\omega^{2}\left(\mathbf{q}\right)}{\omega^{2}-\omega^{2}\left(\mathbf{q}\right)}\propto\\ \notag
 &\iint e^{i\mathbf{q}\cdot\mathbf{r}}e^{-i\mathbf{q}\cdot\mathbf{r'}}V^{\text{ion}}\left(\mathbf{r}-\mathbf{r}'\right)d^{3}rd^{3}r'.
\end{align}
We represent the right-hand side of Eq.\ \eqref{gl24} as a matrix element $\left<m\left|V^{\text{ion}}\left(\mathbf{r}-\mathbf{r'}\right)\right|n\right>$ for the transition between the initial state $\left|n\right>$
to a final state $\left|m\right>$ of a system of electrons and of ions.
Thereby we describe the wavefunction $\Psi\left(\left\{\mathbf{r}_{i}\right\},\left\{\mathbf{R}_{n}\right\}\right)$ for the electrons at positions $\left\{\mathbf{r}_{i}\right\}$
and the ions at positions $\left\{\mathbf{R}_{n}\right\}$ in Born-Oppenheimer approximation,
\begin{align}\label{gl25C}
 \left|\Psi\left(\left\{\mathbf{r}_{i}\right\};\left\{\mathbf{R}_{n}\right\}\right)\right>=\left|\Phi\left(\left\{\mathbf{r}_{i}\right\},\left\{\mathbf{R}_{n}\right\}\right)\right>\left|\chi\left(\left\{\mathbf{R}_{n}\right\}\right)\right>,
\end{align}
where $\left|\Phi\left(\left\{\mathbf{r}_{i}\right\};\left\{\mathbf{R}_{n}\right\}\right)\right>$ is the many electron wavefunction for given momentary positions $\left\{\mathbf{R}_{n}\right\}$.
The two states $\left|n\right>$ and $\left|m\right>$ have the same ionic state $\left|\chi\right>$ (defined by the phonon occupation numbers), i.e., the transition
from $\left|n\right>$ to $\left|m\right>$ is a purely electronic transition.
This transition is done by exciting electrons with wave vectors $\mathbf{k},\mathbf{k}'$ and energies below the Fermi level $\epsilon_{\text{F}}$ to electrons with 
$\mathbf{k}+\mathbf{q},\mathbf{k'}-\mathbf{q}$ and energies above $\epsilon_{\text{F}}$.
For free electrons the initial state $\left|n\right>$ in real-space representation is
\begin{align}\label{gl25}
 n=\left(\frac{1}{\sqrt{V}}\right)^{N}AS\prod\limits_{\mathbf{\widetilde{k}}\neq\mathbf{k},\mathbf{k}'}e^{i\mathbf{\widetilde{k}}\cdot\mathbf{\widetilde{r}}}e^{i\mathbf{k}\cdot\mathbf{r}}e^{i\mathbf{k}'\cdot\mathbf{r}'},
\end{align}
where we have assumed that the spins of the electrons remain unchanged during the excitation so that the spin functions are the same for $\left|n\right>$ and $\left|m\right>$, respectively,
and thus can be omitted.
Of course, in real spin-flip transitions the spins of the electrons are changed.
But this has nothing to do with the formal representation of $V^{ion}_{\mathbf{k},\mathbf{k}'}$ by a transition matrix element.
$AS$ is the antisymmetrization operator.
The final state is
\begin{align}\label{gl26}
 m=\left(\frac{1}{\sqrt{V}}\right)^{N}AS\!\!\prod\limits_{\mathbf{\widetilde{k}}\neq\mathbf{k},\mathbf{k}',\mathbf{k}+\mathbf{q},\mathbf{k}'-\mathbf{q}}
 \!\!\!\!
 e^{i\mathbf{\widetilde{k}}\cdot\mathbf{\widetilde{r}}}e^{i\left(\mathbf{k}+\mathbf{q}\right)\cdot\mathbf{r}}e^{i\left(\mathbf{k}'-\mathbf{q}\right)\cdot\mathbf{r}'}.
\end{align}
Taking into account the orthogonality of exponentials with different wave vectors, the matrix element $\left<m\left|V^{\text{ion}}\left(\mathbf{r}-\mathbf{r}'\right)\right|n\right>$ is given by the right-hand side of Eq.\ \eqref{gl24}, i.e.,
$V^{\text{ion}}_{\mathbf{\widetilde{k}}\mathbf{\widetilde{k}}'}$ indeed can be written as a matrix element for a many-electron system of free electrons.
The idea then is that $V^{\text{ion}}_{\mathbf{k},\mathbf{k}'}$ is for a real material better approximated by a matrix element involving instead of the free-electron states 
$e^{i\mathbf{k}\cdot\mathbf{r}}$, $e^{i\mathbf{k}'\cdot\mathbf{r}'}$, $e^{i\left(\mathbf{k}+\mathbf{q}\right)\cdot\mathbf{r}}$, $e^{i\left(\mathbf{k}'-\mathbf{q}\right)\cdot\mathbf{r}'}$ the single-electron crystal 
states $\left|\mathbf{k}\right>$, $\left|\mathbf{k}'\right>$, $\left|\mathbf{k}+\mathbf{q}\right>$ and $\left|\mathbf{k}'-\mathbf{q}\right>$.

\subsection{Theory of transition matrix elements in crystals}\label{secIId}
In principal, transitions between various states of the system of electrons and phonons are generated by the interactions $V_{\text{inter}}$ which the electrons feel.
We consider the scattering operator of Coulomb interactions $V_{\text{el}}$ between the electrons and electron-phonon interactions $V_{\text{ph}}$,
\begin{align}\label{gl27}
 V_{\text{inter}}=V_{\text{el}}+V_{\text{ph}}.
\end{align}
The electronic parts of the initial state $\left|n\right>$ and the final state $\left|m\right>$ of the transition are eigenstates of the many-electron Hamiltonian
$\widehat{H}=\widehat{H}_{0}+V_{\text{inter}}$. In a mean-field treatment of the electronic states (e.g., density functional theory) part of $V_{\text{el}}$ is already included in $\widehat{H}_{0}$ 
(namely $v_{\text{eff}}^{\text{Kohn-Sham}}$ in density functional theory).
Then $V_{\text{el}}$ in \eqref{gl27} is the small explicit Coulomb rest interaction, and both $V_{\text{el}}$ and $V_{\text{ph}}$ can be treated in a perturbation approach,
thereby considering the eigenstates of $\widehat{H}_{0}$ as unperturbed states.
By this perturbation theory we determine the initial state $\left|n\right>$ and the final state $\left|m\right>$ of the transition from the unperturbed
eigenstates $\left|n^{0}\right>$ and $\left|m^{0}\right>$ of $\widehat{H}_{0}$.
Starting from $\left|m^{0}\right>$, $\left|n^{0}\right>$ and considering as perturbation just $V_{\text{el}}$, we get in first-order perturbation theory the perturbed states
$\left|n^{0}\right>+\left|n_{1}^{\text{el}}\right>$, $\left|m^{0}\right>+\left|m_{1}^{\text{el}}\right>$.
Considering as perturbation just $V_{\text{ph}}$, we get $\left|n^{0}\right>+\left|n_{1}^{\text{ph}}\right>$, $\left|m^{0}\right>+\left|m_{1}^{\text{ph}}\right>$, with
\begin{align}\label{gl28}
 \left|n_{1}^{\text{el}}\right>=\sum_{l\neq n}\left|l_{0}\right>\frac{\left<l_{0}\left|V_{\text{el}}\right|n_{0}\right>}{E_{n_{0}}-E_{l_{0}}},\\
 \label{gl29}
 \left|m_{1}^{\text{el}}\right>=\sum_{l\neq m}\left|l_{0}\right>\frac{\left<l_{0}\left|V_{\text{el}}\right|m_{0}\right>}{E_{m_{0}}-E_{l_{0}}},\\
 \label{gl30}
 \left|n_{1}^{\text{ph}}\right>=\sum_{l\neq n}\left|l_{0}\right>\frac{\left<l_{0}\left|V_{\text{ph}}\right|n_{0}\right>}{E_{n_{0}}-E_{l_{0}}},\\
 \label{gl31}
  \left|m_{1}^{\text{ph}}\right>=\sum_{l\neq m}\left|l_{0}\right>\frac{\left<l_{0}\left|V_{\text{ph}}\right|m_{0}\right>}{E_{m_{0}}-E_{l_{0}}}.
 \end{align}
Here $\left|l_{0}\right>$ are states of the system of electrons and phonons where the electronic parts of them are eigenstates of $\widehat{H}_{0}$.
$E_{n_{0}},E_{m_{0}}$ and $E_{l_{0}}$ are the total energies of the system of electrons and phonons with unperturbed electronic states.
When we have both perturbations $V_{\text{el}}$ and $V_{\text{ph}}$ we get
\begin{align}
 \label{gl32}
 \left|n^{1}\right>=\left|n^{0}\right>+\left|n_{1}^{\text{el}}\right>+\left|n_{1}^{\text{ph}}\right>,\\
 \label{gl33}
 \left|m^{1}\right>=\left|m^{0}\right>+\left|m_{1}^{\text{el}}\right>+\left|m_{1}^{\text{ph}}\right>.
\end{align}
We now calculate $\left<m^{1}\left|V_{\text{el}}+V_{\text{ph}}\right|n^{1}\right>$.
Thereby $V_{\text{ph}}$ increases or decreases the number of phonons (see Sec.\ \ref{secIIe}).
Because phonon states with different phonon numbers are orthogonal,
only those terms in this matrix element are nonzero  for which the bra and  ket belong to the same number of phonons.
E.g., $\left<m_{0}\left|V_{\text{ph}}\right|n_{0}\right>=0$ because $\left|m_{0}\right>$ and $\left|n_{0}\right>$ have the same phonon numbers but
$V_{\text{ph}}\left|n_{0}\right>$ has different phonon numbers.
Another example is
\begin{align}\label{gl34}
 \left<m_{1}^{\text{ph}}\left|V_{\text{el}}\right|n_{0}\right>=\sum_{l\neq m}\frac{\left<m_{0}\left|V_{\text{ph}}\right|l_{0}\right>\left<l_{0}\left|V_{\text{el}}\right|n_{0}\right>}{E^{0}_{m}-E^{0}_{l}}.
 \end{align}

Here $\left<m_{0}\left|V_{\text{ph}}\right|l_{0}\right>$ is nonzero only if $V_{\text{ph}}\left|l_{0}\right>$ has the same phonon numbers
as $\left|m_{0}\right>$, i.e., if $\left|l_{0}\right>$ has different phonon numbers than $\left|m_{0}\right>$ and $\left|n_{0}\right>$.
Because $V_{\text{el}}\left|n_{0}\right>$ has the same phonon numbers as $\left|m_{0}\right>$ and therefore different phonon numbers than $\left|l_{0}\right>$, 
$\left<l_{0}\left|V_{\text{el}}\right|n_{0}\right>=0$ and $\left<m_{1}^{\text{ph}}\left|V_{\text{el}}\right|n_{0}\right>=0$.
Furthermore, we neglect all terms of third order, e.g., $\left<m_{1}^{\text{ph}}\left|V_{\text{ph}}\right|n_{1}^{\text{ph}}\right>$. Finally, the terms $\left<m_{0}\left|V_{\text{el}}\right|n_{1}^{\text{el}}\right>$
and $\left<m_{1}^{\text{el}}\left|V_{\text{el}}\right|n_{0}\right>$ contain only electronic contributions, i.e., they generate $V^{\text{el}}_{\mathbf{k},\mathbf{k}'}$ and we exclude them
because we aim at representing $V^{\text{ion}}_{\mathbf{k},\mathbf{k}'}$ by a transfer matrix element,
and we omit the  term $\left<m_{0}\left|V_{\text{el}}\right|n_{0}\right>$ because it represents the unscreened Coulomb interaction.
It remains
\begin{align}\label{gl35}
 &\left<m^{1}\left|V_{\text{el}}+V_{\text{ph}}\right|n^{1}\right>=\left<m_{0}\left|V_{\text{ph}}\right|n_{1}^{\text{ph}}\right>+\left<m^{\text{ph}}_{1}\left|V_{\text{ph}}\right|n_{0}\right>
\end{align}
Thereby we have used in the last step the fact that $\left<m_{0}\left|V_{\text{ph}}\right|n_{0}\right>$ is zero.
We pause, to give an intuitive physical justification for the use of the perturbation approach.
In conventional perturbation theory it is  assumed that the  perturbation, here $V_{\text{ph}}$, arises because a phonon is present.
However, in our model the initial and the final state have the same phonon numbers.
So what is the physical reason for the perturbation?
The answer is that in a field-theoretical approach a phonon can be spontaneously emitted by  a field fluctuation and this phonon then acts as perturbation.
The phonon then is again absorbed after a very short time.

We thus have shown that $\left<m_{0}\left|V_{\text{ph}}\right|n_{1}^{\text{ph}}\right>+\left<m_{1}^{\text{ph}}\left|V_{\text{ph}}\right|n_{0}\right>$ is indeed a representation of the contribution of ionic screening, $V^{\text{ion}}_{\mathbf{k},\mathbf{k}'}$, 
to the effective electron-electron Coulomb interaction, which is stated (but not proved) already in Ref.\ \onlinecite{8}. 
This result is true although it is a bit counter-intuitive because the matrix element does not contain $V_{\text{el}}$.
\subsection{Simple model for $V_{\text{ph}}$}
\label{secIIe}
The general electron-phonon scattering operator for the scattering of an electron in state $\left|\mathbf{k},j\right>$, where j is the band index,
is in second quantization given by
\begin{align}\label{gl36}
 V_{\text{ph}}=\sum_{\mathbf{k},\mathbf{q},j,j',\lambda}M^{\lambda}_{j\mathbf{k}j'\mathbf{k'}}\left(\widehat{b}_{\mathbf{q}\lambda}+\widehat{b}^{\dagger}_{-\mathbf{q}\lambda}\right)\widehat{c}^{\dagger}_{\mathbf{k}+\mathbf{q},j'}\widehat{c}_{\mathbf{k},j}, 
\end{align}
where $\widehat{b}_{\mathbf{q},\lambda}$ and $\widehat{b}_{-\mathbf{q},\lambda}$ are phonon annihilation and creation operators and where $\widehat{c}^{\dagger}_{\mathbf{k}+\mathbf{q},j}$ and 
$\widehat{c}_{\mathbf{k},j}$ are creation and annihilation operators for electronic states.
In (6.15) of Ref.\ \onlinecite{8} a slightly simpler $V_{\text{ph}}$ is used because it is assumed that interband-scatterings $\left(j\rightarrow j'\right)$ can be neglected.
In contrast to (6.15) we omit the summation over reciprocal lattice vectors $\mathbf{G}$, because we assume that if $\mathbf{k},\mathbf{k}'$ are not in the first Brillouin zone they are,
respectively, brought back into the first Brillouin zone by adding a $\mathbf{G}$.

In the following a simple $V_{\text{ph}}$ is used (as it was done in Ref.\ \onlinecite{8}) for calculating $\left<m_{0}|V_{\text{ph}}|n_{1}^{\text{ph}}\right>$ and $\left<m_{1}^{\text{ph}}\left|V_{\text{ph}}\right|n_{0}\right>$ which considers
only intraband transitions, so that a one-band model suffices (no index $j$), and we neglect a possible dependence of $M$ on $\mathbf{k}$ and $\lambda$,
\begin{align}\label{gl37}
 V_{\text{ph}}=\sum_{\mathbf{k''},\mathbf{q}'}M\left(\mathbf{q}'\right)\left(\widehat{b}_{\mathbf{q}'}+\widehat{b}^{\dagger}_{-\mathbf{q}'}\right)\widehat{c}^{\dagger}_{\mathbf{k}''+\mathbf{q}'}\widehat{c}_{\mathbf{k}''}.
\end{align}
This is of  course an approximation in the phenomenological model, which can be justified only in retrospect
if the resulting phenomenological expression agrees well with ab-initio results.

\subsection{Explicit calculation of the matrix elements}\label{secIIf}

We now calculate by use of Eqs.\ \eqref{gl30}, \eqref{gl31} and of Eq.\ \eqref{gl37} for $V_{\text{ph}}$ the sum of the matrix elements occurring in Eq.\  \eqref{gl35}
\begin{align}\label{gl42}
&\left<m_{0}|V_{\text{ph}}|n_{1}^{\text{ph}}\right>+\left<m_{1}^{\text{ph}}\left|V_{\text{ph}}\right|n_{0}\right>=\\ \notag
&\sum_{l\neq n}\frac{\left<m_{0}\left|V_{\text{ph}}\right|l_{0}\right>\left<l_{0}\left|V_{\text{ph}}\right|n_{0}\right>}{E_{n_{0}}-E_{l_{0}}}+\sum_{l\neq m}\frac{\left<m_{0}\left|V_{\text{ph}}\right|l_{0}\right>\left<l_{0}\left|V_{\text{ph}}\right|n_{0}\right>}{E_{m_{0}}-E_{l_{0}}}
\end{align}
with 
\begin{align}\label{gl43}
 \left|m_{0}\right>=\widehat{c}^{\dagger}_{\mathbf{k}'+\mathbf{q}}\widehat{c}^{\dagger}_{\mathbf{k}-\mathbf{q}}\widehat{c}_{\mathbf{k}}\widehat{c}_{\mathbf{k}'}\left|n_{0}\right>.
\end{align}
As in Ref.\ \onlinecite{8} we consider zero temperature, and this means that the phononic state $\left|n_{0}^{\text{ph}}\right>$ of $\left|n_{0}\right>$ is
\begin{align}\label{gl44}
 \left|n_{0}^{\text{ph}}\right>=\left|0,0,\cdots,0,\cdots\right>.
\end{align}
The matrix elements on the right-hand side of Eq.\ \eqref{gl42} are non-zero only if the respective bra (e.g., $\left<m_{0}\right|$) belongs to the same 
phononic and electronic state as the ket (e.g., $V_{\text{ph}}\left|l_{0}\right>$).
For a given $\mathbf{q}$ this is only the case for
\begin{align}\label{gl45}
 \left|l_{0}\right>=\widehat{b}^{\dagger}_{\mathbf{q}}\widehat{c}^{\dagger}_{\mathbf{k}-\mathbf{q}}\widehat{c}_{\mathbf{q}}\left|n_{0}\right>.
\end{align}
In Ref.\ \onlinecite{8} a second intermediate state $\left|l\right>$ has been considered which, however, belongs to a wavevector $-\mathbf{q}$ and should not be taken
into account when considering the matrix element for given $\mathbf{q}$.
The denominators of Eq.\ \eqref{gl42} are 
\begin{align}\label{gl46}
 E_{n_{0}}-E_{l_{0}}=\epsilon_{\mathbf{k}}-\epsilon_{\mathbf{k}-\mathbf{q}}-\hbar\omega_{\mathbf{q}}
\end{align}
\begin{align}\label{gl47}
 E_{m_{0}}-E_{l_{0}}=-\epsilon_{\mathbf{k}'}-\epsilon_{\mathbf{k}'+\mathbf{q}}-\hbar\omega_{\mathbf{q}}
\end{align}
The matrix elements occurring in \eqref{gl42} are
\begin{align}\label{gl48}
 \left<m_{0}\left|V_{\text{ph}}\right|l_{0}\right>=M\left(\mathbf{q}\right),
\end{align}
\begin{align}\label{gl49}
 \left<l_{0}\left|V_{\text{ph}}\right|n_{0}\right>=M\left(\mathbf{-q}\right)=M^{*}\left(\mathbf{q}\right).
\end{align}
Altogether we have
\begin{align}\label{gl50}
 &\left<m_{0}\left|V_{\text{ph}}\right|n^{\text{ph}}_{1}\right>+\left<m_{1}^{\text{ph}}\left|V_{\text{ph}}\right|n_{0}\right>=V_{\mathbf{k},\mathbf{k}',\mathbf{q}}=\notag\\
&\left|M\left(\mathbf{q}\right)\right|^{2}\left(\frac{1}{\epsilon_{\mathbf{k}}-\epsilon_{\mathbf{k}-\mathbf{q}}-\hbar\omega_{\mathbf{q}}}
+\frac{1}{-\epsilon_{\mathbf{k}'}+\epsilon_{\mathbf{k}'+\mathbf{q}}-\hbar\omega_{\mathbf{q}}}\right).
 \end{align}
Eq.\ \eqref{gl50} formally is the matrix element $V_{\mathbf{k},\mathbf{k}',\mathbf{q}}$ for the transition (originating from the electron-phonon interaction) 
of two electrons in states $\left|\mathbf{k}\right>$ and $\left|\mathbf{k}'\right>$ with energies $\epsilon_{\mathbf{k}}$ and $\epsilon_{\mathbf{k}'}$
below Fermi level $\epsilon_{\text{F}}$ to states $\left|\mathbf{k}-\mathbf{q}\right>$, $\left|\mathbf{k}'+\mathbf{q}\right>$ with energies $\epsilon_{\mathbf{k}-\mathbf{q}}$ and $\epsilon_{\mathbf{k}'+\mathbf{q}}$
 above $\epsilon_{\text{F}}$, and --- again formally --- it looks like an effective electron-electron interaction.
 Thereby $\mathbf{q}$ is completely independent of $\mathbf{k}$ and $\mathbf{k}'$, i.e., the total interaction potential between the electrons in states
 $\left|\mathbf{k}\right>$ and $\left|\mathbf{k}'\right>$ is $\sum_{\mathbf{q}}V_{\mathbf{k},\mathbf{k}'\mathbf{q}}$.
 We now look at the complete potential interaction energy of all electrons in all states $\left|\mathbf{k}\right>$ and $\left|\mathbf{k}'\right>$
 with energies below $\epsilon_{\text{F}}$,
 \begin{align}\label{gl51}
  &V_{\text{total}}= \frac{1}{2}\sum_{\mathbf{k},\mathbf{k}'\mathbf{q}}V_{\mathbf{k},\mathbf{k}'\mathbf{q}}=\notag\\
  &\frac{1}{2}\!\sum_{\mathbf{k},\mathbf{k}'\mathbf{q}}\!\left|M\left(\!\mathbf{q}\right)\right|^{2}\!\left(\frac{1}{\epsilon_{\mathbf{k}}-\epsilon_{\mathbf{k}-\mathbf{q}}-\hbar\omega_{\mathbf{q}}}
\!+\!\frac{1}{-\epsilon_{\mathbf{k}'}+\epsilon_{\mathbf{k}'+\mathbf{q}}-\hbar\omega_{\mathbf{q}}}\right)\!=\!\notag\\
&\sum_{\mathbf{k},\mathbf{k}'\mathbf{q}}\left|M\left(\mathbf{q}\right)\right|^{2}\frac{\hbar\omega_{\mathbf{q}}}{\left(\epsilon_{\mathbf{k}+\mathbf{q}}-\epsilon_{\mathbf{k}}\right)^{2}-\left(\hbar\omega_{\mathbf{q}}\right)^{2}}.
 \end{align}
The factor $\frac{1}{2}$ has been introduced to avoid double counting. 
For the last step thereby a renaming of the summation indices has been performed $\mathbf{k}'\rightarrow\mathbf{k}$, $\mathbf{q}\rightarrow\mathbf{-q}$
and $\omega_{\mathbf{q}}=\omega_{\mathbf{-q}}$ has been used. Altogether, this yields:
\begin{align}\label{gl52}
 V_{\mathbf{k},\mathbf{k}'\mathbf{q}}=\frac{1}{2}\left|M\left(\mathbf{q}\right)\right|^{2}\frac{\hbar\omega_{\mathbf{q}}}{\left(\epsilon_{\mathbf{k}+\mathbf{q}}-\epsilon_{\mathbf{k}}\right)^{2}-\left(\hbar\omega_{\mathbf{q}}\right)^{2}}.
\end{align}
Remembering that $V_{\mathbf{k},\mathbf{k}'\mathbf{q}}$ is a representation of the contribution of ionic screening to the effective electron-electron 
Coulomb interaction, we can equate $V_{\mathbf{k},\mathbf{k}'\mathbf{q}}$ of Eq.\ \eqref{gl52} to $V^{\text{ion}}_{\mathbf{k},\mathbf{k}}$ of Eq. \eqref{gl15},
which gives the final result
\begin{align}\label{gl53}
 \left|M\left(\mathbf{q}\right)\right|^{2}=\frac{1}{V}\frac{4\pi e^{2}}{q^{2}+k_{\text{TF}}^{2}}\frac{1}{2}\hbar\omega_{\mathbf{q}}.
\end{align}
This corresponds to Eq.\ (26.40) of Ref.\ \onlinecite{7} and Eq.\ (6.44) of Ref.\ \onlinecite{8}.

\subsection{Modification of the phenomenological matrix elements: Introduction of their dependence on the spin states of the involved electrons}
\label{secIII}
Eq.\ \eqref{gl2} shows that the real matrix elements do not depend just on the phonon wavevector $\mathbf{q}$ but both on $\mathbf{k}$ and $\mathbf{k}'$
from which $\mathbf{q}=\pm\left(\mathbf{k}'-\mathbf{k}+\mathbf{G}\right)$ can be calculated ($+$ or $-$ sign for phonon absorption or emission, $\mathbf{G}$ is a reciprocal lattice vector - see introduction).
Furthermore, they depend on band indices j and j' whereas for the derivation of Eq.\ \eqref{gl53} it  has been assumed (see Sec. \ref{secIIe})
that there are only intraband transitions ($j=j'$). The matrix elements also depend on the index $\lambda$ which denotes the three polarization vectors 
$\mathbf{p}_{\mathbf{q}\lambda}$ of the phonon, whereas this dependence has been neglected in Eq.\ \eqref{gl53}.
This is of  course a very strong approximation because the electron-phonon scattering operator contains the scalar product between $\mathbf{p}_{\mathbf{q}\lambda}$
and the gradient of the potential $V^{\alpha}$, and this leads for the case of free electrons to a scalar product $\mathbf{q}\cdot\mathbf{p}_{\mathbf{q}\lambda}$
(see Eq.\ (6.14) of Ref.\ \onlinecite{8}).
For high-symmetry wave vectors $\mathbf{q}$  the polarization vectors are longitudinal ($\mathbf{p}_{\mathbf{q}\lambda}\parallel \mathbf{q}$) and transversal ($\mathbf{p}_{\mathbf{q}\lambda}\perp \mathbf{q}$)
--- for arbitrary $\mathbf{q}$ the polarization vectors are pseudo-longitudinal or pseudotransversal.
This clearly demonstrates that the transition matrix element depends strongly on $\mathbf{q}$. Finally, in systems with spin-orbit coupling the electronic
states $\Psi_{j\mathbf{k}}$ are no pure spin states but spin-mixed states according to \cite{10}
\begin{align}\label{gl54}
 \Psi_{j\mathbf{k}}=\left[a_{j\mathbf{k}}\left(\mathbf{r}\right)\left|\uparrow\right>+b_{j\mathbf{k}}\left(\mathbf{r}\right)\left|\downarrow\right>\right]\exp\left(i\mathbf{k}\mathbf{r}\right),
\end{align}
where $a_{j\mathbf{k}}\left(\mathbf{r}\right)$ and $b_{j\mathbf{k}}\left(\mathbf{r}\right)$ are lattice periodic functions and $\left|\uparrow\right>$, 
$\left|\downarrow\right>$ are the two spinor eigenfunctions of $\widehat{S}_{z}$.
The wave function is denoted as "dominant spin up" or "dominant spin down" if $\left|a_{j\mathbf{k}}\right|^{2}=\int \left|a_{j\mathbf{k}}\left(\mathbf{r}\right)\right|^{2}d^{3}r$
is larger or smaller than $\left|b_{j\mathbf{k}}\right|^{2}=\int \left|b_{j\mathbf{k}}\left(\mathbf{r}\right)\right|^{2}d^{3}r$.
Usually one denotes the dominant spin character by $\widetilde{m}_{s}$, and this index (which is determined by $j\mathbf{k}$) is added to the wave function,
$\Psi_{j\mathbf{k}}^{\widetilde{m}_{s}}$.
In ab-initio theories of electron phonon scatterings one distinguishes between spin flip transitions if in $\Psi^{\widetilde{m}_{s}}_{j\mathbf{k}}$ and 
$\Psi^{\widetilde{m}_{s}'}_{j'\mathbf{k}'}$ we have $\widetilde{m}_{s}\neq \widetilde{m}_{s}'$
and non-spin-flip transitions for $\widetilde{m}_{s}= \widetilde{m}_{s}'$.
In Eq.\ \eqref{gl53} the dominant spin character is not considered,
i.e., it gives the same result independent of the question whether the considered electron-phonon
interaction is a spin-flip  transition or a non-spin-flip transition.
In many materials, however, there is a big difference between these two types of transitions,
because spin mixing is small for most $\Psi_{j\mathbf{k}}$.
To estimate the difference between the two types of transitions we assume that we can write approximately
\begin{align}\label{gl55}
 \Psi^{\widetilde{m}_{s}}_{j\mathbf{k}}\left(\mathbf{r}\right)&=\varphi_{j\mathbf{k}}\left(\mathbf{r}\right)\left[a^{\widetilde{m}_{s}}\left|\uparrow\right>+b^{\widetilde{m}_{s}}\left|\downarrow\right>\right]\exp\left(i\mathbf{k}\mathbf{r}\right)\\ \notag
 &=\varphi_{j\mathbf{k}}\left(\mathbf{r}\right)\left|\chi^{\widetilde{m}_{s}}\right>\exp\left(i\mathbf{k}\cdot\mathbf{r}\right),
 \end{align}
with
\begin{align}\label{gl56}
 &\left|\chi^{\text{dominant up}}\right>=\begin{pmatrix} \sqrt{1-b^{2}}\\b \end{pmatrix},\\ \notag
 &\left|\chi^{\text{dominant down}}\right>=\begin{pmatrix}
                                         b\\\sqrt{1-b^{2}}
                                        \end{pmatrix}.
\end{align}
Thereby $b^{2}$ is the averaged spin-mixing factor calculated by the average \eqref{gl53} $\left<\left|b_{j\mathbf{k}}\right|^{2}\right>$
of $b_{j\mathbf{k}}^{2}$ over all states involved in the electron-phonon scattering processes \cite{14}.
Therefore the square of  the transition matrix element is for $\widetilde{m}_{s}\neq \widetilde{m}_{s}'$
\begin{align}\label{gl57}
 \left|\left<\Psi^{\widetilde{m}_{s}'}_{j'\mathbf{k}'}\left|V_{\text{ph}}\right|\Psi^{\widetilde{m}_{s}}_{j\mathbf{k}}\right>\right|^{2}&=\left|\left[2b\sqrt{1-b^{2}}\left<\varphi_{j'\mathbf{k}'}\left|V_{\text{ph}}\right|\varphi_{j\mathbf{k}}\right>\right]\right|^{2}\\ \notag
 &\approx 4 b^{2}\left|\left<\varphi_{j'\mathbf{k}'}\left|V_{\text{ph}}\right|\varphi_{j\mathbf{k}}\right>\right|^{2},
\end{align}
whereas for $\widetilde{m}_{s}=\widetilde{m}_{s}'$ we have
\begin{align}\label{gl58}
 \left|\left<\Psi^{\widetilde{m}_{s}'}_{j'\mathbf{k}'}\left|V_{\text{ph}}\right|\Psi^{\widetilde{m}_{s}}_{j\mathbf{k}}\right>\right|^{2}=\left|\left<\varphi_{j'\mathbf{k}'}\left|V_{\text{ph}}\right|\varphi_{j\mathbf{k}}\right>\right|^{2},
\end{align}
i.e, the squares of spin-flip transition matrix elements are typically a factor of $4b^{2}$ smaller than the squares of  non-spin-flip transition matrix elements.
For Ni the ab-initio calculated value \cite{14} $b^{2}=0.025$.

\section{Conclusions}\label{secIV}
Transition matrix elements for electron-phonon scatterings are important for many processes in solid state physics.
The electron-phonon interaction has two contributions, one arising from the  phonon-induced distortion of the lattice potential (Elliott part)
and one resulting from the phonon-induced distortion of the  spin-orbit coupling (Yafet part).
In the present paper the Elliott part is considered.

In the literature the transition-matrix elements are calculated in two ways.
There is an approximate expression for the matrix elements derived (for example in the textbooks of Ashcroft and Mermin \cite{7} or of Czycholl \cite{8})
by the combination of a phenomenological theory of electronic and ionic screening of the electron-electron interaction with a microscopic
perturbation theory for the matrix elements.
In other papers the matrix elements are calculated by the ab-initio electron theory.
A comparison shows that the matrix elements calculated in these two ways differ very strongly \cite{N14}, 
which in principle does not justify to use the
approximate expression. However, it is shown \cite{N14} that possibly it can be used for the calculation of
macroscopic observables which involve a weighted summation of matrix elements. 
This is shown \cite{N14} for the example of the demagnetization rate of a ferromagnetic film after excitation with a fs laser pulse.
Unfortunately, however, this does not mean that the approximated matrix element can be used also for a calculation of other macroscopic observables
for which they enter in another weighted way.

As a future project we will consider the transition matrix elements for the Yafet part of the electron-phonon interaction.
In the literature there are already microscopic treatments of the electronic screening of the Yafet electron-phonon interaction
by the spin-other-orbit interaction \cite{5} or by the spin-same-orbit interaction \cite{6}.
In Ref.\ \onlinecite{6} a variety of contributions to the electronic screening of the Yafet part are found,
however, no explicit results are given.
We will try to figure out how large the various contributions are.
Finally, we will investigate whether there are also ionic screening effects for the Yafet part of the electron-phonon interaction. 
 
\bibliography{NScreening}
\bibliographystyle{apsrev4-1}
\end{document}